\documentclass[12pt]{article}
\newcommand{\text}{\rm}
\begin{document}
\title{Dual description of $U(1)$ charged fields\\in $(2+1)$
  dimensions}
\author{C.~D.~Fosco$^a$, V.~E.~R.~Lemes$^b$, L.~E.~Oxman$^b$\\
  S.~P.~Sorella$^b$ and O.~S.~Ventura$^b$\\ \\
  {\normalsize\it $^a$Centro At\' omico Bariloche, 8400 Bariloche,
    Argentina}\\ \\
  {\normalsize\it $^b$Departamento de F\'{\i}sica Te\'{o}rica, Instituto de F\'{\i}sica,}\\
  {\normalsize\it Universidade do Estado do Rio de Janeiro}\\
  {\normalsize\it Rua S{\~a}o Francisco Xavier, 524, 20550-013, Maracan\~{a}}\\
  {\normalsize\it Rio de Janeiro, Brazil}\\
  {\normalsize\it UERJ/DFT-02/2000} }
\date{\today}
\maketitle
\begin{abstract}
  We show that the functional bosonization procedure can be
  generalized in such a way that, to any field theory with a conserved
  Abelian charge in $2+1$ dimensions, there corresponds a dual Abelian
  gauge field theory.  The properties of this mapping and of the dual
  theory are discussed in detail, presenting different explicit
  examples. In particular, the meaning and effect of the coefficient
  of the Chern-Simons term in the dual action is interpreted in terms
  of the spin and statistics connection.
\end{abstract}
\bigskip \renewcommand{\theequation}{\thesection.\arabic{equation}}
\section{Introduction}
Recently, the functional bosonization of fermionic systems in $(2+1)$
dimensions has been a subject of intense research, with various
attempts to extend to higher dimensions some of the non perturbative
techniques and results known for the two-dimensional case~\cite{2b}.
These efforts have led to a set of interesting and promising
results~\cite{result1}-\cite{result6}, which turned out to be
relevant, for instance, to the understanding of the universal behavior
of the Hall conductance in interacting electron systems~\cite{bo}. An
important point to be remarked is that, in this approach, the
bosonization of a charged fermionic system in \mbox{$(2+1)$}
dimensions is achieved through the introduction of an Abelian vector
gauge field $A_\mu$~\cite{result1}-\cite{result7}.

More precisely, as in the two-dimensional case, the $U(1)$ fermionic
current~\mbox{$J_\mu =\bar\psi \gamma_\mu \psi$} is mapped into a topologically
conserved current $j_\mu^T$~\cite{result1}-\cite{result6}:
\begin{equation}
 \label{bos-curr}
J_\mu =\bar{\psi}\gamma_\mu \psi \,\longrightarrow \,j_\mu^T
=\epsilon_{\mu \nu \lambda}\partial _\nu A_\lambda \;.
\end{equation}
Correspondingly, the free massive Dirac action
\begin{equation}
S_F[\bar{\psi },\psi]=\int d^3 x\, \bar\psi \left( \partial \!\!\!/+m \right) \psi \ ,
\label{faction}
\end{equation}
is transformed into a bosonic, gauge invariant action $S_B[A]$
\begin{equation}
  \label{act-bos}
S_F[\bar{\psi },\psi]\longrightarrow S_B[A]\;,
\end{equation}
whose exact form is, in general, unknown, since its evaluation
requires the calculation of a fermionic determinant in $(2+1)$
dimensions. It is possible, however, to carry out the bosonization
procedure in such a way that it is consistent with a perturbative
expansion of the fermionic determinant, by a decoupling transformation
of the fermionic fields in the functional
integral~\cite{Fosco:1998gm}.

However, just on general grounds, we may say that $S_B$
consists of a Chern-Simons action (the leading term), plus an infinite
series of terms depending on the curvature $\widetilde{F}_\mu
=\frac{1}{2}\epsilon_{\mu\nu\lambda} \partial_\nu A_\lambda$, namely
\begin{equation}
S_B[A]\;=\; i \,\frac{1}{\eta} \,S_{CS}[A]\;+\; R [\widetilde{F}]\;,
\label{act-bos1}
\end{equation}
where $S_{CS}[A]$ stands for the Chern-Simons action~\footnote{We
  adopt Euclidean conventions.}
\begin{equation}
 \label{c-s}
S_{CS}[A]\;=\;\frac{1}{2}\int d^3x \, \epsilon_{\mu\nu\lambda}A_\mu
\partial _\nu A_\lambda \;,
\end{equation}
and $R [\widetilde{F}]$ denotes the remainder, higher order
contributions in the curvature, including those which are non local and
non quadratic in $\widetilde{F}$~\cite{result1}-\cite{result6}.  It is
worth mentioning that, recently, the current bosonization rule
(\ref{bos-curr}) has been proven to hold even in the presence of
interactions, giving to eq.(\ref{bos-curr}) a universal character
\cite{result6}. This means that the bosonization rule for a fermionic
system whose action contains an interaction part $I[J^\mu]$, depending
only on the current $J^\mu$, generalizes to:
\begin{equation}
S_F[,{\bar\psi},\psi]+I[J^\mu] \leftrightarrow S_B[A]+I[\epsilon^{\mu
\nu \rho}\partial_\nu A_\rho]
\label{urules}
\end{equation}
where $S_B[A]$ is the same functional that bosonizes the {\it free}
fermionic action $S_F[\psi]$. This relationship has to be understood as
an equality between correlation functions of currents, as obtained from
the action corresponding to each description, namely
\begin{equation}
\langle J_{\mu_1} ... J_{\mu_n} \rangle_{(S_F+I[J])} = \langle
j^T_{\mu_1} ... j^T_{\mu_n} \rangle_{(S_B+I[j^T])} \ .
\label{univ-rules}
\end{equation}

The parameter $\eta$ in the leading term of eq.(\ref{act-bos}) is the
coefficient of the induced Chern-Simons action. It turns out that the
determination of this coefficient is plagued by finite ambiguities,
related to the choice of the regularization procedure. In other words,
as usual in any field theory computation, a set of physical conditions
has to be imposed on the system in order to fix all the ambiguities. We
shall, for the time being, leave this parameter unspecified, coming
back to it later on, when discussing the relationship of $\eta$ with
the statistics of the excitations present in the bosonized version of
the theory.

In the low energy regime, only the Chern-Simons term survives in the
right hand side of eq.$\left( {\rm{\ref{act-bos}}}\right) $, and this
yields a simple, closed expression for the bosonized action:
\begin{equation}
 \label{m-inf}
\lim_{m\to \infty }S_B[A]\;=\; i \,\frac{1}{\eta} \,S_{CS}[A]\;.
\end{equation}

The aim of this work is to generalize this construction, originally
used for the bosonization of fermionic actions, to the case of an
arbitrary field theory model with a conserved global $U(1)$ charge. We
shall see that, regardless of the spin (and statistics) of the fields
in the original action, any three-dimensional model with an Abelian
conserved charge can be mapped into a dual Abelian gauge theory.  We
shall refer to this mapping as {\em duality\/} rather than {\em
bosonization}, in view of the more general meaning of its defining
characteristics.  As the most distinctive feature of this mapping, we
mention that the conserved Noether current $J_\mu$ corresponding to the
$U(1)$ global symmetry will, again, have a topological current
$j_\mu^T$ as its dual partner in the Abelian gauge theory:
\begin{equation}
\label{u1-curr}
\,J_\mu \longrightarrow \;j_\mu^T =\epsilon_{\mu \nu \lambda}\partial_\nu A_\lambda \;.
\end{equation}
The topological current $j_\mu^T$ becomes then identified with ${\tilde
  F}_\mu$, the {\em topological} {\em local}, {\em gauge invariant}, and
{\em identically conserved\/} current that can be written in terms of
$A_\mu$ in the dual gauge invariant theory. Moreover, the gauge
invariance of the bosonized theory will be shown to be related to the
renormalizability of the functional integrals leading to the dual
theory.

Another property of the bosonization procedure, which will be extended
to the general case, shall be that the dual gauge invariant action
will also contain a Chern-Simons term, whenever there is a breaking of
parity in the original theory.

\noindent It is worth noting  that the possibility of describing
three-dimensional $U(1)$ charged models in terms of gauge fields can
bring us new perspectives in order to understand in a deeper way the
non perturbative dynamics of planar systems. Needless to say, much is
already known about the geometry and the topology of gauge theories.

The present work is organized as follows: in section~\ref{dirac} we
briefly review the bosonization of the massive Dirac field, fixing our
conventions, and extracting from that case the essential properties of
the procedure.  This procedure is then generalized to an arbitrary
field with a $U(1)$ charge in section~\ref{general}.  In
section~\ref{examples} a few examples are dealt with in detail.
Section~\ref{inter} is devoted to the physical interpretation of the
coefficient of the induced Chern-Simons in the dual gauge invariant
action form. Finally, in section~\ref{concl}, we present our
conclusions.

\section{Massive Dirac field}\label{dirac}
Let us begin by defining ${\mathcal Z}[s_\mu]$, the generating
functional of current correlation functions for a massive Dirac field
in the presence of an external source $s_\mu$, in its
`fermionic representation':
\begin{equation}
  \label{eq:defzf}
  {\mathcal Z}[s_\mu] \;=\; \int {\mathcal D}{\bar\psi} {\mathcal
    D}\psi \, \exp\{- S_F[{\bar\psi},\psi; s_\mu]\} \;,
\end{equation}
where
\begin{equation}
S_F[{\bar\psi},\psi; s_\mu] \;=\;\int d^3x \, {\bar\psi}(x)[\not\!\partial + m +
    i \not \! s (x)] \psi(x) \;,
\end{equation}

and we adopted the Euclidean spacetime conventions:

\begin{equation}
  \label{eq:conv1}
  g_{\mu\nu}\,=\, \delta_{\mu\nu} \;\;\; (\gamma_\mu)^\dagger \,=\,
  \gamma_\mu \;\;\; \{ \gamma_\mu , \gamma_\nu \}\,=\, 2\,
  \delta_{\mu\nu} \;.
\end{equation}

Current correlation functions are averages of the conserved current
operator $J_\mu (x) = {\bar\psi}(x)\gamma_\mu \psi(x)$
$$
\langle J_{\mu_1} (x_1) \ldots J_{\mu_n} (x_n) \rangle_s \;=\; \frac{1}{{\mathcal
    Z}[s_\mu]} \int {\mathcal D}{\bar \psi} {\mathcal D}\psi \, J_{\mu_1} (x_1) \ldots
J_{\mu_n} (x_n)
$$
\begin{equation}
 \label{eq:defcf}
\times \exp\{-\int d^3x {\bar\psi}[\not\!\partial + m + i \not \! s]
\psi\} \;,
\end{equation}
and may, of course, be obtained by functional differentiation,
$$
\langle J_{\mu_1} (x_1) \ldots J_{\mu_n} (x_n) \rangle_s \;=\;
\frac{1}{{\mathcal Z}[s_\mu]}
i \frac{\delta}{\delta s_{\mu_1}(x_1)} \ldots i
\frac{\delta}{\delta s_{\mu_n}(x_n)}{\mathcal Z}[s_\mu]
$$

\begin{equation}
  \label{eq:funcdif}
=\; i \frac{\delta}{\delta s_{\mu_1}(x_1)} \ldots i\frac{\delta}{\delta s_{\mu_n}(x_n)}
{\mathcal W}[s_\mu] \;,
\end{equation}
with ${\mathcal W}[s_\mu]\;=\; \ln {\mathcal Z}[s_\mu]$. Correlation
functions for vanishing $s_\mu$ shall be denoted as in (\ref{eq:defcf}),
except for the omission of the subindex `$s$' in the average symbol.

The first step in the bosonization procedure consists of performing
the non-anomalous change of variables:
\begin{equation}
  \label{eq:stp1}
  \psi(x) \;\to\; e^{i \alpha (x)} \psi(x) \;\;\;\;\;\; {\bar\psi}(x)
  \;\to\; e^{-i \alpha (x)} {\bar\psi}(x)
\end{equation}
for the fermionic fields in the functional integral. This leads to an
equivalent expression for ${\mathcal Z}[s_\mu]$,
\begin{equation}
  \label{eq:stp2}
 {\mathcal Z}[s_\mu] \;=\; \int {\mathcal D}{\bar\psi} {\mathcal
    D}\psi \, \exp \left\{- S_F[{\bar\psi},\psi; s_\mu + \partial_\mu \alpha] \right\} \;,
\end{equation}
or,
\begin{equation}
  \label{eq:stp2a}
 {\mathcal Z}[s_\mu] \;=\; {\mathcal Z}[s_\mu + \partial_\mu \alpha] \;.
\end{equation}

As the left hand side of (\ref{eq:stp2a}) is independent of $\alpha$,
so must be the right hand side. Then, we obtain an equivalent form for
${\mathcal Z}[s_\mu]$ by integrating over $\alpha$ and, if we want to
keep also source-independent factors, dividing by the corresponding
volume factor ${\mathcal N}_1$:
$$
{\mathcal Z}[s_\mu] \;=\;\frac{1}{{\mathcal N}_1}\, \int {\mathcal
  D}\alpha \; {\mathcal Z}[s_\mu + \partial_\mu\alpha]
$$
 \begin{equation}
  \label{eq:stp3}
  \;=\;\frac{1}{{\mathcal N}_1}\, \int {\mathcal
    D}\alpha  {\mathcal D}{\bar\psi} {\mathcal D}\psi \,
    \exp \left\{- S_F[{\bar\psi},\psi; s_\mu + \partial_\mu \alpha] \right\} \;,
\end{equation}
where ${\mathcal N}_1 = \int{\mathcal D}\alpha$. This is of course a divergent
factor, similar to the gauge group volume element which is factorized
in the application of the Faddeev-Popov procedure to an Abelian gauge
invariant theory. We assume that it has been regulated, for example,
by introducing an Euclidean cutoff $\Lambda$, and by using a finite volume
spatial region. We shall not need its explicit form, however, and will
keep the same notation ${\mathcal N}_1$, in the understanding that
this object has been regularized.  Then we go on to an expression
where the integration over the scalar field $\alpha$ is transformed into
one over a `pure gauge' vector field $b_\mu$, such that $b_\mu = \partial_\mu \alpha$.
The integration over this vector field is thus constrained to satisfy
a null-curvature condition, namely
\begin{equation}
  \label{eq:nullc}
  {\mathcal Z}[s_\mu] \;=\;\frac{{\mathcal N}_2}{{\mathcal N}_1}\, \int {\mathcal
    D}b_\mu  \delta[{\tilde f}_\mu(b)]{\mathcal D}{\bar\psi} {\mathcal D}\psi \,
    \exp \left\{- S_F[{\bar\psi},\psi; s_\mu + b_\mu] \right\} \;,
\end{equation}
where ${\tilde f}_\mu\,=\,\epsilon_{\mu\nu\lambda}\partial_\nu b_\lambda$. In (\ref{eq:nullc}), the new
constant factor ${\mathcal N}_2$ appears as a consequence of the fact
that the $\delta$ function of ${\tilde f}_\mu (b)$ differs from the `pure
gauge' condition for $b_\mu$ by a constant, ill-defined factor.  One
easily sees that:
\begin{equation}
  \label{eq:defn2}
  {\mathcal N}_2 \;=\; \det (\epsilon_{\mu\lambda\nu}
  \partial_\lambda) \;.
\end{equation}
This factor is actually zero, because of the zero eigenvalue
corresponding to the longitudinal mode of the operator $\epsilon_{\mu\lambda\nu} \partial_\lambda$,
and it is formally canceled by an analogous factor in the denominator,
which comes from the $\delta$ function. A possible way to cure this would
be to represent the determinant as a functional integral over vector
Grassmann fields,
\begin{equation}
  \label{eq:n21}
   {\mathcal N}_2 \;=\; \int {\mathcal D}{\bar c}_\mu {\mathcal
     D}c_\mu \; \exp\{\int d^3x {\bar c}_\mu \epsilon_{\mu\lambda\nu} \partial_\lambda c_\nu \}
\end{equation}
and to fix the gauge (through the Faddeev-Popov procedure) for this
gauge invariant functional integral. We shall not detail this,
however, since we will not need the explicit form of ${\mathcal N}_2$.
We shall keep the notation `${\mathcal N}_2$' to denote this (gauge
fixed) object.

One then shifts the $b_\mu$ field: $b_\mu \to b_\mu - s_\mu$ to decouple it from
the source $s_\mu$ in the fermionic action, and exponentiates the $\delta$
function by using a Lagrange multiplier field $A_\mu$, to obtain
\begin{equation}
  \label{eq:stp4}
  {\mathcal Z}[s_\mu] \,=\,\frac{{\mathcal N}_2}{{\mathcal N}_1}\,
  \int {\mathcal D}A_\mu {\mathcal D}b_\mu {\mathcal D}{\bar\psi}
  {\mathcal D}\psi \,
    \exp \left\{- S_F[{\bar\psi},\psi; b_\mu ] + i \int d^3x A_\mu
    {\tilde f}_\mu(b - s ) \right\} \,.
\end{equation}
Equation (\ref{eq:stp4}) then leads to the bosonic representation
\begin{equation}
 \label{eq:stp5}
  {\mathcal Z}[s_\mu] \;=\; \int {\mathcal D}A_\mu \,
e^{-S_B[A] - i \int d^3x s_\mu \epsilon_{\mu\nu\lambda}\partial_\nu A_\lambda}
\end{equation}
where the `bosonized action' $S_B[A]$ is defined by
\begin{equation}
  \label{eq:stp6}
e^{- S_B[A]} \;=\; \frac{{\mathcal N}_2}{{\mathcal N}_1}
 \int {\mathcal D} b_\mu \,\det[\not\!\partial + m + i \not \! b ]\,
 e^{i\int d^3 x b_\mu \epsilon_{\mu\nu\lambda} \partial_\nu
   A_\lambda} \;.
\end{equation}

The role of the factors ${\mathcal N}_{1,2}$ is evident in
(\ref{eq:stp6}): their presence adds a constant to the bosonized
action. To understand their effect, one can consider the $s_\mu =0$
case, which leads to
$$
{\mathcal Z}[0] \;=\; \int {\mathcal D}A_\mu \, e^{-S_B[A]}
$$
\begin{equation}
 \label{eq:s0}
\;=\;  \frac{{\mathcal N}_2}{{\mathcal N}_1}
   \int {\mathcal D}A_\mu  \int {\mathcal D} b_\mu \,
  \det[\not\!\partial + m + i \not \! b ]\,
 e^{i\int d^3 x b_\mu \epsilon_{\mu\nu\lambda} \partial_\nu  A_\lambda} \;,
\end{equation}
or, integrating over $A_\mu$,
$$
{\mathcal Z}[0] \;=\; \frac{{\mathcal N}_2}{{\mathcal N}_1} \int
{\mathcal D} b_\mu \,\delta[{\tilde f}(b)] \, \det[\not\!\partial + m
+ i \not \! b ] \;,
$$
$$
=\; \frac{1}{{\mathcal N}_1} \int {\mathcal D}\alpha \,
\det[\not\!\partial + m + i \not \! \partial \alpha ] \;=\;
\frac{1}{{\mathcal N}_1} \int {\mathcal D}\alpha \,
\det[\not\!\partial + m ]
$$
\begin{equation}
 \;=\; \det[\not\!\partial + m ] \;,
\end{equation}
where we used the gauge invariance of the fermionic determinant. This
shows that the factors are required if one is interested in evaluating
quantities that are interesting even in the absence of sources, like
the vacuum energy in a gravitational background, or in a finite
volume. In what follows, however, we shall disregard the constant
factors, since we will be mostly interested in flat spacetime
correlation functions, where those factors cancel out.

To find the explicit form of the bosonized action, one needs of course
to evaluate the fermionic determinant {\em and\/} of the integral over
the auxiliary field $b_\mu$, which yields an expression with the
structure of (\ref{act-bos1}). To make sense of expression
(\ref{eq:stp6}) when the fermionic determinant is evaluated beyond the
quadratic approximation, one should say something about the
renormalizability of the functional integral over the auxiliary field
$b_\mu$. We note that this integral could be regarded as corresponding
to the generating functional of complete Green's functions for a
dynamical vector gauge field $b_\mu$, equipped with a gauge invariant
`action':
\begin{equation}
  \label{eq:defsb}
   S_b[b_\mu]\;=\; - {\mathcal W}[b_\mu] \;,
\end{equation}
and with a `source' $s^{b}_\mu\,=\,i \epsilon_{\mu\nu\lambda}\partial_\nu
A_\lambda$, namely,
\begin{equation}
  \label{eq:fintb}
  e^{-S_B[A]} \;=\; \int [ {\mathcal D}b_\mu ]\, e^{- S_b[b_\mu] + \int
    d^3x  s^{b}_\mu b_\mu} \;,
\end{equation}
where $[{\mathcal D}b_\mu]$ denotes the measure including gauge
fixing. We assume the Landau gauge has been chosen, for the sake of
simplicity.

The UV properties of this integral are not explicit, due to the
non-local and non-polynomial character of $S_b$. However, by using a
functional expansion of ${\mathcal W}$, we see that
$$
S_b[b_\mu]\;=\; - \frac{1}{2}\int d^3x_1 d^3x_2 {\mathcal
    W}^{(2)}_{\mu_1\mu_2}(x_1,x_2) b_{\mu_1}(x_1) b_{\mu_2}(x_2)
$$
\begin{equation}
 \label{eq:fexp}
\,-\,\frac{1}{3!}\int d^3x_1 d^3x_2 d^3x_3 {\mathcal
     W}^{(3)}_{\mu_1\mu_2\mu_3}(x_1,x_2,x_3) b_{\mu_1}(x_1) b_{\mu_2}(x_2)
   b_{\mu_3}(x_3) \,+\, \ldots
\end{equation}
where we adopted the notation:
\begin{equation}
  \label{eq:defdw}
  {\mathcal W}^{(n)}_{\mu_1\cdots\mu_n}(x_1,\cdots,x_n)\;=\;
 \left. \frac{\delta^n{\mathcal W}[b_\mu]}{\delta b_{\mu_1}(x_1)
     \cdots \delta b_{\mu_n}(x_n)}\right|_{b_\mu = 0} \;.
\end{equation}

The first step to understand the convergence properties of the
functional integral over $b_\mu$, is to find out the large momentum
behaviour of the propagator, which is obviously determined by
${\mathcal W}^{(2)}$. Regardless of the details of the theory, the
Fourier transform of the function ${\mathcal W}^{(2)}$ will grow
linearly with $k$, for large values of the momentum $k$, as a
dimensional analysis shows: gauge invariance (together with the Landau
gauge choice) of the fermionic determinant implies that ${\mathcal
  W}^{(2)}$ is transverse, and its coefficient has to be a scalar
function of $k_\mu$, which for large values of $|k|$ behave like $\sim
|k|$.  Then the $b_\mu$ field propagator will behave like $\sim 1/k$
in the same regime.  We note that, in the absence of gauge invariance
for the fermionic determinant, there is no reason to exclude worse
behaviours for the propagator.

The $b_\mu$ propagator will connect to generalized vertices
corresponding to all the ${\mathcal W}^{(n)}$'s. These vertices are
momentum dependent, and actually, they behave as a negative power of
$k$ for large $k$.  For an arbitrary $n$, we have the behaviour
${\mathcal W}^{(n)} \sim k^{3-n}$, with $n$ even and greater than $4$.
Thus the power counting of a 1PI diagram $G$ corresponding to the
functional integral over $b_\mu$ will have a superficial degree of
divergence $\omega(G)$ given by
\begin{equation}
  \label{eq:sdiv}
  \omega (G)\;=\; 3 L - I + \sum_{n \geq 4} k_n (3-n) \;,
\end{equation}
where $L$ is the number of independent loops, $I$ the number of
internal $b_\mu$ lines, and $k_n$ the number of vertices of $n$ legs
in the diagram $G$. Using the standard relation $L = I - V +1$, where
$V$ is the total number of vertices (of any kind), we see that
\begin{equation}
  \label{eq:cons}
  \omega (G)\;=\; 3 + 2 I - 3 V  + \sum_{n \geq 4} k_n (3-n) \;.
\end{equation}
On the other hand, any proper diagram verifies the `topological'
relation:
\begin{equation}
  \label{eq:topo}
  \sum_{n \geq 4} n k_n \;=\; 2 I + E \;,
\end{equation}
where $E$ denotes the number of external $b_\mu$ field lines.  When
(\ref{eq:topo}) is inserted into (\ref{eq:cons}), it yields
\begin{equation}
  \label{eq:sdivf}
  \omega (G)\;=\; 3 - E \;,
\end{equation}
which evidently corresponds to a renormalizable theory, since only the
two-point function {\em might\/} require a subtraction.  Gauge
invariance, combined with the linear degree of divergence, imply that
the counterterm could only be a Chern-Simons like term. This, however, is
not renormalized beyond one loop~\cite{coleman}.

We conclude
from this discussion that the bosonized action exists in perturbation
theory, to all orders (in a loop expansion). It should be noted that,
when the fermionic determinant is evaluated in a derivative expansion,
the previous power counting is spoiled. This is, however, just an
artifact of the approximation.

To conclude this section, we enumerate some essential features of the
bosonization algorithm, to take them into account in the
generalization we shall develop in the next section:
\begin{enumerate}

\item The starting point should be a model with a {\em global\/}
  $U(1)$ invariance. A local invariance is not good, since then the
  original action is not changed by (the analog of) the coordinate
  transformation (\ref{eq:stp1}).
\item The generating functional for the current correlation functions,
  ${\mathcal Z}[s_\mu]$, should verify (\ref{eq:stp2a}). This allows for
  the decoupling of the source $s_\mu$ and the $b_\mu$ field through a
  shift of the latter. Of course, the way to prove that this equation
  is verified will, in general, depend on the particular model
  considered, and it may, as in the fermionic case, require a change
  of variables for the charged fields.
\item As a matter of principle, we should require the bosonized action
  to exist, and to avoid defining it through a non-renormalizable
  functional integral. We have seen that, if the property
  (\ref{eq:stp2a}) holds, that integral is finite. Moreover, there is
  no need to assume that $ S_b=- {\mathcal W}[b_\mu] $ proceeds from a
  fermionic matter field; just gauge invariance of the result is
  required.
\end{enumerate}
\section{The general framework}\label{general}
As mentioned at the the end of the preceeding section, the extension
of the procedure originally developed for the Dirac field to an
arbitrary model with a global Abelian charge is straightforward, as
long as property (\ref{eq:stp2a}) holds. The reason is that then all
the steps leading to the bosonic representation (\ref{eq:stp5}) hold
true, with the only change that the expression for ${\mathcal Z}[s_\mu]$
as an integral over charged fields is not written explicitly.  One
sees that
\begin{equation}
 \label{eq:gen1}
  {\mathcal Z}[s_\mu] \;=\; \int {\mathcal D}A_\mu \,
e^{-S_B[A] - i \int d^3x s_\mu \epsilon_{\mu\nu\lambda}\partial_\nu A_\lambda}
\end{equation}
where now the `bosonized action' $S_B[A]$ is defined by
\begin{equation}
  \label{eq:gen2}
e^{- S_B[A]} \;=\; \frac{{\mathcal N}_2}{{\mathcal N}_1}
 \int {\mathcal D} b_\mu \,{\mathcal Z}[b_\mu] \,
 e^{i\int d^3 x b_\mu \epsilon_{\mu\nu\lambda} \partial_\nu
   A_\lambda} \;,
\end{equation}
an expression of which (\ref{eq:stp6}) is a particular case, where
${\mathcal Z}[s_\mu]$ is a fermionic determinant. Different field
theories coupled to an external source $s_\mu$ will give rise to an
identical expression, although the representation of ${\mathcal
  Z}[s_\mu]$ as a functional integral over the original matter field
will of course be different for each model. By expanding
(\ref{eq:stp2a}) in powers of $\alpha$ and integrating by parts, one
can see that it is equivalent to the (infinite) set of Ward
identities:
\begin{equation}
  \label{eq:wi}
  \langle \partial \cdot J(x_1) \cdots \partial \cdot J(x_n) \,
J_{\nu_1}(y_1) \cdots J_{\nu_k} (y_k) \rangle_s \;=\;0 \;\;\;, \forall n=1,2,
\ldots, k=0,1,\ldots
\end{equation}
This of course also implies the corresponding identities for $s_\mu =
0$, but one cannot obtain the $s_\mu \neq 0$ identities from the $s_\mu
= 0$ ones. If one just requires that the $s_\mu = 0$ identities hold,
and the source $s_\mu$ is only an artifact to derive $s_\mu = 0$
Green's functions, then there is no lost of generality in assuming
(\ref{eq:stp2a}) (this will of course not affect the $s_\mu=0$
averages). That property is tantamount to requiring that ${\mathcal
  Z}[s_\mu]$ is invariant under `gauge transformations' of the source.
This can be guaranteed if the coupling of the source to the matter
fields is gauge invariant, since then the matter fields can be gauge
transformed in order to compensate for the gauge transformation of the
`gauge field' $s_\mu$. In what follows, we shall then require current
conservation even in the presence of a non-vanishing source $s_\mu$,
as this implies conservation when $s_\mu \to 0$ as well. More
importantly, the gauge invariance is necessary to assure the
finiteness of the $b_\mu$ functional integral.
In order to work out the general setup, now in terms of the matter
fields,  let us consider a pair $%
\phi ,\phi ^{\dagger }$ of complex $U(1)$-charged fields (scalars, vectors,
spinors, \ldots), and let
\begin{equation}
  \label{ac}
S[\phi^\dagger, \phi]\;=\;\int d^3x \,{\cal L}(\phi^\dagger, \phi ;
\partial \phi^\dagger, \partial \phi )\;\;,
\end{equation}
be the corresponding classical action. As already remarked, $S[\phi^\dagger,
\phi]$ will be assumed to be invariant under the global $U(1)$
transformations
\begin{eqnarray}
\phi &\to &e^{i\alpha }\phi \;,  \label{u1tr} \\
\phi ^{\dagger } &\to &e^{-i\alpha }\phi ^{\dagger }\;,  \nonumber
\end{eqnarray}
yielding a conserved $U(1)$ Noether current $J_\mu $, {\it i.e.},
\begin{equation}
 \label{u1curr}
\partial _\mu J_\mu =0
\end{equation}
when the equations of motion hold. Now we require the conservation of
the current even when the source does not vanish. As we said above,
the coupling to the source has to be then gauge invariant, and this is
assured by replacing normal derivatives by covariant derivatives with
respect to the source. Note that, in general, the conserved current
will then depend on the source $s_\mu$.  With this in mind, we introduce
the generating functional ${\mathcal Z}[s_\mu]$
\begin{equation}
  \label{g-f}
{\mathcal Z}[s_\mu]\;=\; \int D\phi D\phi ^{\dagger }e^{-\int d^3x\left( {\mathcal L}%
[\phi^\dagger,\phi;(D\phi)^\dagger,D\phi]\; \right)} \ ,
\end{equation}
where $D_\mu\phi=(\partial_{\mu}+is_\mu) \phi$.  Following the steps already applied in
section~\ref{dirac}, one performs the local change of variables in
(\ref{g-f})
\begin{eqnarray}
\phi (x) &\to &e^{i\alpha (x)}\phi (x)\;,  \label{u1-loc} \\
\phi ^{\dagger }(x) &\to &e^{-i\alpha (x)}\phi ^{\dagger }(x)\;.
\nonumber
\end{eqnarray}
which immediately leads to the desired property
\begin{equation}
  \label{eq:des}
  {\mathcal Z}[s_\mu] \;=\; {\mathcal Z}[s_\mu + \partial_\mu \alpha] \;.
\end{equation}
Using now the following relation
\begin{eqnarray}
{\cal L}\left( \phi ,\phi ^{\dagger },(\partial _\mu  +is_\mu
)\phi ,(\partial _\mu  - is_\mu )\phi ^{\dagger }\right) &=&
\nonumber \\
{\cal L}\left( \phi ,\phi ^{\dagger },\partial _\mu  \phi
,\partial _\mu \phi ^{\dagger }\right) +s_\mu J^\mu \;\;, &&
\label{curr-rel}
\end{eqnarray}
where $J_\mu (\phi ,\phi ^{\dagger })\;$ is the Noether current associated to the
global $U(1)$ invariance of the action $S(\phi )$, for the generating
functional ${\mathcal Z}[s_\mu]$ we obtain
\begin{equation}
  \label{g-f1}
{\mathcal Z}[s_\mu]\;=\; \int D\phi D\phi ^{\dagger }e^{-\int d^3x\left( {\mathcal L}%
[\phi^\dagger,\phi;(\partial\phi)^\dagger,\partial\phi] + s_\mu J^\mu \; \right)} \ .
\end{equation}
The meaning of the equation (\ref{curr-rel}) is that coupling the $U(1)$ current$J^\mu $ to an external source $s_\mu ,$ is equivalent to
replace the space-time derivative $\partial _\mu $ with the covariant derivative ($%
\partial _\mu +is_\mu )$, according to the principle of the minimal coupling.
Although the relation $\left( {\rm {\ref{curr-rel}}}\right) $ is
self-evident for actions which are linear in the space-time
derivatives of the fields, as for instance in the case of the usual
fermionic action, it is worth underlining that eq.$\left( {\rm
    {\ref{curr-rel}}}\right) $ can be in fact satisfied also in more
general cases, the recipe here being that of using a first-order
formalism by means of the introduction of suitable auxiliary fields,
as it will be shown in detail in the next section in the case of the
complex scalar fields.

\noindent Using now eq.(\ref{u1-loc}), we get
$$
{\cal Z}[s_\mu]=\int D\phi D\phi ^{\dagger }D\eta _\mu DA_\mu $$
$$
\exp \left\{-\int d^3x [ {\cal L}( \phi ,\phi ^{\dagger }, (\partial _\mu +i\eta _\mu -is_\mu )\phi
  ,(\partial _\mu -i\eta _\mu +is_\mu ) \phi ^\dagger ]\right.
$$
\begin{equation}
\left. + i \;A_\mu \epsilon ^{\mu \nu \rho }\partial _\nu \eta_\rho ] \right\}\;,
\end{equation}
so that, performing the change of variables
\begin{equation}
\label{change}
\eta _\mu \to \eta _\mu - s_\mu \;,
\end{equation}
the external source $s_\mu $ decouples from the fields $\phi ,\phi
^{\dagger },$ {\it i.e.}
$$
{\cal Z}[s_\mu]=\int D\phi D\phi ^{\dagger }D\eta _\mu DA_\mu \exp \left\{-\int d^3x [ {\cal
    L}( \phi ,\phi ^{\dagger },(\partial _\mu +i\eta_\mu)\phi ,(\partial_\mu -i\eta_\mu)\phi ^{\dagger }]\right.
$$
\begin{equation}
\label{g-f3}
\left. + i \;A_\mu \epsilon ^{\mu \nu \rho }\partial _\nu (\eta_\rho - s_\rho )] \right\}\;,
\end{equation}
Introducing thus the effective action $S_{{\rm eff}}(\eta )=\int d^3x{\cal L}%
_{{\rm eff}}(\eta )\;$through the equation
\begin{equation}
e^{-S_{{\rm eff}}[\eta_\mu ]}=\int D\phi D\phi ^{\dagger }e^{-\int d^3x{\cal L}%
\left( \phi ,\phi ^{\dagger },(\partial _\mu +i\eta _\mu )\phi ,(\partial
_\mu -i\eta _\mu )\phi ^{\dagger }\right) \;\;},  \label{eff-act}
\end{equation}
we have
\[
{\cal Z}(s)=\int D\eta _\mu DA_\mu e^{-\int d^3x\left( {\cal L}_{%
      {\rm eff}}(\eta )\;-i\;A_\mu \epsilon ^{\mu \nu \rho }\partial
    _\nu \eta _\rho + i A_\mu \epsilon ^{\mu \nu \rho }\partial _\nu
    s_\rho \right) \;}.
\]
Finally, defining the so called dual action $S_{{\rm dual}}(A)$
\begin{equation}
  \label{dual-act}
e^{-S_{{\rm dual}}[A_\mu]}=\int D\eta _\mu e^{-\int d^3x\left( {\cal L}_{{\rm eff%
}}(\eta )\;-\;iA_\mu \epsilon^{\mu \nu \rho }\partial _\nu \eta _\rho
\right) \;},\;
\end{equation}
we obtain the {\it dual representation\/} of the generating functional
${\cal Z}[s_\mu]$, namely
\begin{equation}
 \label{g-ff}
{\cal Z}[s_\mu]=\int DA_\mu e^{-\left( S_{{\rm dual}}[A_\mu]+i \int d^3xs_\mu
\epsilon ^{\mu \nu \rho }\partial _\nu A_\rho \right) \;}.
\end{equation}
It should be observed that the dual field $A_\mu $ can be actually
interpreted as a genuine gauge field, as the dual action $S_{{\rm
    dual}}(A)$ in eq.$\left( {\rm {\ref{dual-act}}}\right) $ is
manifestly gauge invariant,
\begin{equation}
S_{{\rm dual}}[A_\mu ]\;=\;S_{{\rm dual}}[A_\mu +\partial _\mu \omega ]\;.
\label{gin}
\end{equation}

Expression (\ref{g-ff}) provides a representation of the
correlation functions in terms of a $%
(2+1) $ gauge field theory. In particular, from the coupling of the
external source $s_\mu $ in the expression $\left( {\rm
    {\ref{g-ff}}}\right) $, it becomes apparent that, as announced,
the$\;U(1)$ Noether current $J_\mu $ is mapped into a dual topological
current
\begin{equation}
\label{dual1}
J_\mu (\phi ,\phi ^{\dagger })\longrightarrow j_\mu^T (A)=\epsilon^{\mu \nu \rho }\partial _\nu A_\rho \;.
\end{equation}
Correspondingly, for the classical action we have
\begin{equation}
\label{dual2}
S[\phi]=\int d^3x {\cal L}\left( \phi ,\phi ^{\dagger },\partial _\mu  \phi
,\partial _\mu \phi ^{\dagger }\right)\longrightarrow S_{{\rm dual}}[A_\mu]\;.
\end{equation}
Equations $\left( {\rm {\ref{dual1}}}\right) ,\;\left( {\rm {\ref{dual2}}}%
\right) \;$represent our duality mapping and have to be understood in
terms of an equality among the correlations functions,
$$
\left\langle J_{\mu _1}(x_1)J_{\mu
    _2}(x_2)\;\ldots\;J_{\mu _n}(x_n)\right\rangle _{S[\phi]}
$$
\begin{equation}
 \label{corr}
=\;\left\langle j_{\mu _1}^T(x_1)j_{\mu _2}^T(x_2)\;\ldots\;
j_{\mu_n}^T(x_n)\right\rangle _{S_{{\rm dual}}[A_\mu]}\;
\end{equation}
obtained by differentiating the generating functionals $\left( {\rm
    {\ref {g-f}}}\right) $ and $\left( {\rm {\ref{g-ff}}}\right) $
with respect to the external source $s_\mu $. Equation $\left( {\rm
    {\ref{corr}}}\right) $ states that the Green's functions of $U(1)$
Noether currents have a dual representation in terms of topological
currents built up with a gauge field.

As a final comment we remark that the universal character of the
bosonization rules for fermionic systems (eq.(\ref{urules})), are
generalized to the present dual mapping. This means that, adding to
the initial action $S[\phi]$ a current interaction term $I[J^\mu]$, the
final dual result is found to be
\begin{equation}
S[\phi]+I[J^\mu] \leftrightarrow S_{{\rm dual}}[A]+I[\epsilon^{\mu \nu \rho}\partial_\nu A_\rho] \ .
\label{udual}
\end{equation}
Indeed, this can be easily seen by representing the interaction term
$I[J^\mu]$ in the following Fourier functional integral form
\begin{equation}
e^{-I[J^\mu]}=\int Da_\mu e^{-(\int d^3x a_\mu J^\mu + \tilde{I}[a_\mu])}
\label{rep}
\end{equation}
for a suitable action $\tilde{I}[a_\mu]$. Therefore, for the generating
functional we get
\begin{eqnarray}
\lefteqn{{\mathcal Z}[s_\mu]\;=\; \int D\phi D\phi ^{\dagger } e^{-(S[\phi] +I[J^\mu]+
\int d^3x s_\mu J^\mu)}\nonumber }\\
&&=\int D\phi D\phi ^{\dagger }Da_\mu e^{-(S[\phi] +\tilde{I}[a_\mu]+
\int d^3x (s_\mu+a_\mu) J^\mu)} \ .
\end{eqnarray}
Performing now the path integration over the variables $\phi$, $\phi^\dagger$
using the dual representation (\ref{g-ff}), it follows
\begin{equation}
{\mathcal Z}[s_\mu]=\int DA_\mu Da_\mu e^{-(S_{{\rm dual}}[A_\mu] +\tilde{I}[a_\mu]+
\int d^3x (s^\mu+a^\mu)j^T_\mu) } \ ,
\end{equation}
which, upon integration over $a_\mu$ and use of eq.(\ref{rep}), yields
the final dual representation
\begin{equation}
{\mathcal Z}[s_\mu]=\int DA_\mu  e^{-(S_{{\rm dual}}[A_\mu] + I[j^T_\mu]+
\int d^3x s^\mu j^T_\mu) } \ ,
\end{equation}
which implies, in particular, eq.(\ref{udual}).

\section{Examples}\label{examples}
For a better understanding of the previous results, let us work out in
detail some examples, beginning by the simplest case of the dual
mapping applied to a complex scalar field.
\subsection{Complex scalar field}
The model is described by the $U(1)$ invariant action
\begin{equation}
S(\varphi ,\varphi ^{\dagger })=\int d^3x\left( \partial _\mu \varphi
\partial ^\mu \varphi ^{\dagger }+m^2\varphi \varphi ^{\dagger }\;+V(\left|
\varphi \right| )\right) \;,  \label{sc-act}
\end{equation}
$\varphi ,\;\varphi ^{\dagger }\;$being complex scalar fields and $V(\left| \varphi \right| )$
stands for the scalar potential. As already underlined, it is
convenient to switch to the equivalent first order formalism by
introducing suitable auxiliary fields. In the present case this task
amounts to rewriting the action $\left( {\rm {\ref{sc-act}}}\right) $
in the following form:
\begin{equation}
S(\varphi ,b)=\int d^3x\left( b_\mu ^{\dagger }\partial ^\mu \varphi +b_\mu
\partial ^\mu \varphi ^{\dagger }-b_\mu ^{\dagger }b^\mu +m^2\varphi \varphi
^{\dagger }\;+V(\left| \varphi \right| )\right) \;.  \label{s-a}
\end{equation}
Expression $\left( {\rm {\ref{s-a}}}\right) $ is easily seen to be
completely equivalent to $\left( {\rm {\ref{sc-act}}}\right) $ upon
elimination of the two auxiliary fields $b_\mu ,\;b_\mu ^{\dagger }$
through the equations of motion
\begin{eqnarray}
\frac{\delta S}{\delta b^\mu } &=&\partial ^\mu \varphi ^{\dagger }-b_\mu^{\dagger }\;,  \label{eq-m} \\
\frac{\delta S}{\delta b_\mu ^{\dagger }} &=&\partial ^\mu \varphi -b^\mu \;.
\nonumber
\end{eqnarray}
The Noether current $J^\mu $ corresponding to the $U(1)$ global
invariance
\begin{eqnarray}
\varphi &\to &e^{i\alpha }\varphi \;,\;\;\;\;\;\;\;\;\;\;\;b_\mu
\to e^{-i\alpha }b_\mu \;  \label{u1-inv} \\
\varphi ^{\dagger } &\to &e^{-i\alpha }\varphi ^{\dagger
}\;,\;\;\;\;\;\;\;\;\;b_\mu ^{\dagger }\to e^{i\alpha }b_\mu
^{\dagger }\;,  \nonumber
\end{eqnarray}
is found to be
\begin{eqnarray}
J^\mu &=&i\left( b^{\dagger \mu }\;\varphi -b^\mu \varphi ^{\dagger
}\right) \;,  \label{u1-sc} \\
\partial _\mu J^\mu &=&0\;+\;{\rm eqs.\;of\;motion\;.}  \nonumber
\end{eqnarray}
For the generating functional ${\cal Z}(s)$
\begin{equation}
{\cal Z}(s)={\cal N}\int D\varphi D\varphi ^{\dagger }Db_\mu Db_\mu^{\dagger }e^{-\left( S(\varphi ,b)\;+\;\int d^3xs_\mu J^\mu \right)
\;},
\label{sc-g-f}
\end{equation}
we get
$$
{\cal Z} \;=\;{\cal N}\int D\varphi D\varphi ^{\dagger }Db_\mu
Db_\mu ^{\dagger }D\eta _\mu DA_\mu
$$
\begin{equation}
e^{-\left( S(\varphi ,(\partial +i\eta )\varphi ,b)+\int d^3x(s_\mu
J^\mu +\;A_\mu \epsilon ^{\mu \nu \rho }\partial _\nu \eta _\rho
)\right) \;}.
\label{sc-g-f1}
\end{equation}
Moreover, according to eq.$\left( {\rm {\ref{curr-rel}}}\right) ,$ the
following relationship is easily proven to hold
\begin{equation}
S(\varphi ,(\partial +i\eta )\varphi ,b)+\int d^3xs_\mu J^\mu
=S(\varphi ,(\partial +i(\eta +s)\varphi ,b)\;.
\label{sc-curr-rel}
\end{equation}
Repeating now the same steps of the previous section, for the final
form of the generating functional ${\cal Z}(s)\;$we obtain
\begin{equation}
{\cal Z}(s)={\cal N}\int DA_\mu e^{-\left( S_{{\rm dual}}(A)+\int d^3xs_\mu
\epsilon ^{\mu \nu \rho }\partial _\nu A_\rho \right) \;},  \label{sc-gf}
\end{equation}
with the gauge invariant dual action $S_{{\rm dual}}(A)$ being defined
by

\begin{equation}
e^{-S_{{\rm dual}}(A)\;}=\int D\varphi D\varphi ^{\dagger }Db_\mu Db_\mu^{\dagger }D\eta _\mu e^{-\left( S(\varphi ,(\partial +i\eta )\varphi
,b)+\int d^3xA_\mu \epsilon ^{\mu \nu \rho }\partial _\nu \eta _\rho
\right) \;}\;.  \label{sc-d-act}
\end{equation}

We see therefore that, as announced, the Noether current $\left( {\rm
    {\ref {u1-sc}}}\right) $ is mapped into the topological current \
$J_T^\mu (A)=\epsilon ^{\mu \nu \rho }\partial _\nu A_\rho $.
Equation $\left( {\rm {\ref{sc-gf}}}\right) $ defines the dual
representation of a charged scalar field in terms of a $(2+1)$ gauge
theory. It is also worth mentioning that, in the present case, some of
the one-loop contributions to the dual gauge invariant action $S_{{\rm
    dual}}(A)$ have been computed \cite {chaichian}. In particular,
from the expression of the vacuum polarization
\cite{chaichian}, it follows that there is no induced Chern-Simons term in $%
S_{{\rm dual}}(A)$. As expected, this result is due to the absence of
parity breaking terms in the starting action $\left( {\rm
    {\ref{sc-act}}}\right)$.
\subsection{Complex vector field}
As a second example, let us discuss the case of the complex vector
field whose classical action reads
\begin{equation}
S(B,B^{\dagger })=\int d^3x\left( \epsilon ^{\mu \nu \rho }B_\mu^{\dagger }\partial _\nu B_\rho -\mu B_\mu ^{\dagger }B^\mu \right) \;.
\label{B-act}
\end{equation}

This model is known as the complex self-dual model \cite{self-dual},
and it has been proven to be equivalent to the (complex)
Maxwell-Chern-Simons model. It describes a massive excitation of mass
$\mu$ and spin $s=\mu /|\mu |$. It should also be observed that the
action $\left( {\rm {\ref {B-act}}}\right) $ is not parity invariant,
due to the presence of the mass term $\mu B_\mu ^{\dagger }B^\mu $.

\noindent Obviously, expression $\left( {\rm {\ref{B-act}}}\right) $ is left
invariant by the global $U(1)$ transformations

\begin{eqnarray}
B_\mu &\to &e^{i\alpha }B_\mu  \label{B-u1} \\
B_\mu ^{\dagger } &\to &e^{-i\alpha }B_\mu ^{\dagger }\;,  \nonumber
\end{eqnarray}
which yield the following Noether current $J^\mu $
\begin{eqnarray}
J^\mu (B,B^{\dagger }) &=&i\epsilon ^{\mu \nu \rho }B_\nu^{\dagger }B_\rho \;,  \label{B-curr} \\
\partial _\mu J^\mu &=&0\;+\;{\rm eqs.\;of\;motion\;.}  \nonumber
\end{eqnarray}
As usual, let us start with the generating functional ${\cal Z}(s)$
\begin{equation}
{\cal Z}(s)={\cal N}\int DBDB^{\dagger }e^{-\left( S(B,B^{\dagger
})\;+\;\int d^3xs_\mu J^\mu \right) \;}.  \label{B-func}
\end{equation}
Repeating the same steps as before, it can be easily shown that the
dual form for the generating functional $\left( {\rm
    {\ref{B-func}}}\right) $ is found to be
\begin{equation}
{\cal Z}(s)={\cal N}\int DA_\mu e^{-\left( S_{{\rm dual}}(A)+\int d^3xs_\mu
\epsilon ^{\mu \nu \rho }\partial _\nu A_\rho \right) \;},  \label{B-dual}
\end{equation}
with
\begin{equation}
e^{-S_{{\rm dual}}(A)\;}=\int DB_\mu DB_\mu ^{\dagger }D\eta _\mu e^{-\left(
S(B,(\partial +i\eta )B)+\int d^3xA_\mu \epsilon ^{\mu \nu \rho }\partial_\nu \eta _\rho \right) \;}\;,  \label{B-dual-act}
\end{equation}
being the corresponding gauge invariant dual action. We see thus that,
once again, the $U(1)$ Noether current $J^\mu $ is mapped into
the topological current $J_T^\mu (A)$
\begin{equation}
J^\mu (B,B^{\dagger })=i\epsilon ^{\mu \nu \rho }B_\nu ^{\dagger}B_\rho \;\longrightarrow J_T^\mu (A)=\epsilon ^{\mu \nu \rho }\partial_\nu A_\rho \;.
\label{Bm-curr}
\end{equation}
Concerning now the form of the dual gauge invariant action $S_{\rm dual}(A)$,
  the presence of an induced Chern-Simons term can be easily
  confirmed, due to the parity breaking mass term in the starting
  action (\ref{B-act}). This statement follows also by simple
  inspection of the Feynman rules needed for the computation of the
  effective dual action $S_{\rm dual} (A)$. Indeed, from the
  expression~(\ref{B-dual-act}) we see that to each interaction vertex
  of the type $BB^{\dagger }\eta$ there is an associated factor $i\epsilon _{\mu \nu \rho},$ namely
\begin{equation}
\left[ (BB^{\dagger }\eta )-{\rm vertex}\right] {\rm {\ }\longrightarrow
i\epsilon _{\mu \nu \rho }\;,}  \label{vertex}
\end{equation}
Moreover, for the propagator $\left\langle B_\mu ^{\dagger }B_\nu \right\rangle $ we have
\begin{eqnarray}
\left\langle B_\mu ^{\dagger }B_\nu \right\rangle &=&\;{\cal G}_{\mu \nu }^{%
{\rm even}}(p)\;+\;{\cal G}_{\mu \nu }^{{\rm odd}}(p)\;,  \label{B-prop} \\
{\cal G}_{\mu \nu }^{{\rm even}}(p) &=&\frac \mu {p^2-\mu ^2}\left( g_{\mu
\nu }-\frac{p_\mu p_\nu }{\mu ^2}\;\right) \\
{\cal G}_{\mu \nu }^{{\rm odd}}(p) &=&\frac{i\epsilon _{\mu \nu \lambda
}p^\lambda }{p^2-\mu ^2}\;,
\end{eqnarray}
${\cal G}_{\mu \nu }^{{\rm even}}(p),\;{\cal G}_{\mu \nu }^{{\rm odd}}(p)$
denoting the even and the odd contributions, respectively. Looking
then at the one-loop contribution for the vacuum polarization diagram,
it is easily realized that the combination of vertices and propagators
will always lead to an odd parity contribution, coming from
expressions of the type
\begin{equation}
\epsilon ^{\mu \alpha \beta }\epsilon ^{\nu \sigma \tau }\int \frac{%
d^3k}{(2\pi )^3}{\cal G}_{\alpha \tau }^{{\rm odd}}(p-k){\cal G}_{\beta
\sigma }^{{\rm even}}(k)\;.  \label{pol}
\end{equation}
These contributions, properly renormalized, will result in the
existence of a Chern-Simons term in the dual effective action
$S_{{\rm dual}}(A)$.

\noindent Let us end this section by spending a few words on the interesting
case in which the parity violating mass term is absent in the initial
action. We start therefore with a topological action, corresponding to
a complex pure Chern-Simons term, {\it i.e.}
\begin{equation}
S_{{\rm top}}=\int d^3x\epsilon ^{\mu \nu \rho }B_\mu ^{\dagger
}\partial _\nu B_\rho \;.
\label{c-s-t}
\end{equation}
Taking into account that the $U(1)$ Noether current $\left( {\rm {\ref
      {B-curr}}}\right) $ gets unmodified, we end up with a functional
generator whose dual expression is provided by the same formula
$\left( {\rm {\ref {B-dual}}}\right) $, with $S_{{\rm dual}}(A)$ given
by
\begin{equation}
e^{-S_{{\rm dual}}(A)\;}=\int DB_\mu DB_\mu ^{\dagger }D\eta _\mu e^{-\left(
S_{{\rm top}}(B,(\partial +i\eta )B)+\int d^3xA_\mu \epsilon ^{\mu \nu
\rho }\partial _\nu \eta _\rho \right) \;}\;.  \label{d-c-s}
\end{equation}
The only difference with the previous massive case relies on the form
of the propagator, which now reads
\begin{equation}
\left\langle B_\mu ^{\dagger }B_\nu \right\rangle _{{\rm top}}=\frac{%
i\epsilon _{\mu \nu \lambda }p^\lambda }{p^2}\;.  \label{c-s-prop}
\end{equation}
We could expect thus that, in spite of the presence of the $\epsilon_{\mu \nu
  \rho }$ tensor in the starting action (\ref{c-s-t}), no induced
Chern-Simons term should be generated in the final effective dual action
$S_{{\rm dual}}(A)$. This should follow form the observation that  a
generic Feynman diagram obtained from the action $S_{\rm top }\left(
  B,(\partial +i\eta )B\right)$ will contain as many vertices as propagators, so
that there will be no $\epsilon$ tensor left at the end. However, as in the massless fermionic case {\cite{large}}, the use of a gauge invariant regularization ({\it i.e.} Pauli-Villars) will introduce a parity even term, which no longer protects the generation of a possible induced Chern-Simons term in the effective action. 
\subsection{Fractional statistics model}
We may consider as a starting point a theory of Dirac fermions coupled
to a statistical gauge field $a_\mu$, with a partition function
$$
{\mathcal Z}[s_\mu]\;=\; \int {\mathcal D}a_\mu {\mathcal D}{\bar\psi} {\mathcal D}\psi \, \exp \left\{- \int d^3x{\bar\psi}  [\not \! \partial + i \not \! a + i \not \! s + m ] \psi \right.
$$
\begin{equation}
  \label{eq:fest1}
\left. + i \theta
S_{CS}[a_\mu] \right\} \;.
\end{equation}
It should be noted that the statistical field $a_\mu$ is not related
to $A_\mu$, the field appearing in the dual representation, but is
introduced just as a trick to represent the statistical interaction.
This representation is of course simpler than the one that would
result from using `anyon field operators' and no statistical gauge
field.  As the Dirac field describes fermions, we see that the total
statistical phase under particle interchange is $e^{i \pi \nu}$,
where:
\begin{equation}
  \label{eq:totest}
  \nu \;=\; 1 \,+\, \frac{1}{2 \pi \theta} \;.
\label{estat}
\end{equation}

Let us now see what is the dual theory that corresponds to this model,
by a direct application of the procedure described in
Sect.~\ref{general}.
We first note that the partition function is, indeed, invariant under
gauge transformations of $s_\mu$, and that $a_\mu$ must be integrated,
since it is really part of the original matter (anyonic) field.

Then, integrating over $a_\mu$, we obtain
\begin{eqnarray}
\lefteqn{  {\mathcal Z}[s_\mu]\;=\; \int  {\mathcal
    D}{\bar\psi} {\mathcal D}\psi \, \exp \left\{- \int d^3x{\bar\psi}
[\not \! \partial +  i \not \! s + m ] \psi \right. + \nonumber}\\
&&\left. +\frac{i}{\theta}\; \int d^3x d^3y J_\mu(x) {\mathcal K}_{\mu\nu}(x-y) J_\nu(y)
\right\}
\label{int-curr}
\end{eqnarray}
where $J_\mu ={\bar\psi} \gamma_\mu \psi$ is the fermionic current and ${\mathcal
  K}_{\mu\nu}$ is defined by
\begin{equation}
\epsilon_{\mu\lambda\rho} \partial^x_\lambda K_{\rho\nu}(x-y)
\;=\;\delta_{\mu\nu} \delta^{(3)} (x-y) \;.
\label{1}
\end{equation}
We are left thus with the bosonization of a massive Dirac fermion in
the presence of an interaction depending only on the current $J_\mu(x)$.
Therefore, making use of the universality properties of the
bosonization rules for the fermionic current given in
eq.(\ref{urules}) (see also ref.\cite{result6}), the above expression
(\ref{int-curr}) is bosonized into
\begin{equation}
S_B[A]\;=\; i \,\frac{1}{\eta} \,S_{CS}[A]\;+\;R[\widetilde{F}] +  \frac{i}{\theta}\int d^3x d^3y j_\mu^T(x) {\mathcal K}_{\mu\nu}(x-y) j_\nu^T(y)
\;,
\label{stat-bos}
\end{equation}
with $j_\mu^T = \epsilon_{\mu\nu\lambda} \partial_{\nu}A_{\lambda} $ being the topological current.
Moreover, from eq. (\ref{1}), we easily obtain
\begin{equation}
S_B[A]\;=\; i \,(\frac{1}{\eta} + \frac{1}{\theta})\,S_{CS}[A]\;+\;R[\widetilde{F}]
\;.
\label{stat-fin}
\end{equation}
We are now ready to discuss the physical interpretation of the results
so far obtained.
This will be the task of the next section.


\section{Physical interpretation}\label{inter}
Let us now deal with the dual gauge invariant action for the field
$A_\mu$, with the aim of interpreting some of its properties.

We shall see that the general structure of the dual action is strongly
related to the statistical properties of its excitations.

The bosonized action will certainly display nontrivial excitations
corresponding to field configurations with finite energy, labeled by
an integer $n$, corresponding to the quantization of the `magnetic'
flux {\cite{kov}}. Indeed, this quantization is a consequence of the presence of
the $R$-term in eq.(\ref{stat-fin}), which implies that a finite
energy field configuration $A_\mu$ must be a (regular) pure gauge at
spatial infinity. As a matter of fact, the energy density is entirely
due to the $R$-term, whose leading low energy contribution is found to
be the Maxwell term. These configurations correspond typically to
localized vortices.

If we now consider a configuration in the dual theory where two well
separated vortices are adiabatically interchanged, the final phase
factor relative to the case where the vortices are not interchanged
should be dominated by the Chern-Simons term in eq.(\ref{stat-fin}),
which can be re-expressed as
\begin{equation}
i(\frac{1}{\eta}+\frac{1}{\theta})\int d^3x d^3y j_\mu^T(x) {\mathcal K}_{\mu\nu}(x-y) j_\nu^T(y)~.
\label{fase}
\end{equation}
In addition, taking into account that $\partial_\mu j_\mu^T(x)=0$, we see
that the term (\ref{fase}) corresponds to the well-known form of the
`statistical interaction' between particle currents (see, for
example,~\cite{lerda}). This means that under an interchange of the
positions of two vortices in the dual gauge theory, the wave function
will pick up a phase factor $e^{i \pi \nu}$, where
\begin{equation}
  \label{eq:stat}
  \nu \;= \frac{1}{2\pi}(\frac{1}{\eta}+\frac{1}{\theta}).
\end{equation}

At this point, in order to go further in the analysis of the
statistics of these excitations, we have to specify the value of the
induced Chern-Simons coefficient $\eta$. As already underlined in the
introduction, the computation of this coefficient is plagued by finite
regularization ambiguities, which have to be fixed by imposing
suitable physical requirements on the theory. In what follows, we
shall adopt a determination of $\eta$ compatible with the requirement of
invariance of the three dimensional fermionic determinant under large
gauge transformations, when the euclidean time coordinate is
compactified to a circle $S^1$\cite{large}.

In particular, for the free massive Dirac fermion, this condition
leads to the value
\begin{equation}
\eta \;=\; \frac{1}{2\pi}\;.
\end{equation}

We underline here that with this value of $\eta$, the statistical factor
in eq.(\ref{eq:stat}) coincides precisely with that of
eq.(\ref{eq:totest}).

It is also interesting to understand this result from the spin and
statistics theorem point of view. If one assumes that the density of
particles (defined in terms of the time component of the bosonized
current) is localized, and that point-like densities are avoided by
the introduction of smeared-out densities (that tend to point-like
ones at the end of the calculation), one can show~\cite{lerda} that
the spin of each excitation becomes
\begin{equation}
  \label{eq:spin}
  S \;=\; \frac{1}{2} \nu \;.
\end{equation}
This is the well known result that the spin and statistics theorem
holds when the interaction is mediated by a Chern-Simons gauge field.
However, note that here we see an interesting new relation: the
statistics (and spin) of the matter field is encoded in the
coefficient of the induced Chern-Simons term. Of course, the same
analysis holds also in the case where the initial fractional
statistical action is that of a spinless scalar complex field, in the
presence of the statistical field $a$. In this case the statistical
factor $\nu$ contains only the contribution coming from the $\theta$ term,
due to the absence of the induced Chern-Simons term in the bosonic
polarization tensor.

\section{Conclusions}\label{concl}
We have shown that the bosonization program in three dimensions can be
generalized to any Abelian field theory system displaying a global
U(1) invariance. In particular, the $U(1)$ Noether current $J_\mu$ is
mapped into the topological dual current $j^T_\mu$, this mapping being
exact and universal.

We would like to observe that, in the case where a second order field
theory is considered, the above mentioned properties of the dual
mapping have been derived by recasting the theory in a first order
formalism. Moreover, the current correlation functions corresponding
to these two formalisms are known to differ by contact terms, i.e.\ by
terms proportional to the derivatives of the delta function. However,
these terms are not relevant to the calculation of correlation
functions at different points, implying thus the vanishing of the
contact terms.  Also as a by-product of the use of a first-order
formulation, we note that the decoupling transformations
of~\cite{Fosco:1998gm} can be trivially extended to the case of
arbitrary matter fields, by replacing the Dirac operator with the
appropriate one. This would produce the formal relation between
interacting ($\Psi$) and free ($\chi$) matter fields in the functional
integral, which generalizes the one for the Dirac case:
\begin{eqnarray}
\Psi (x) &=& [1 + e (\not \! \partial + m)^{-1}
\not \!\! A]^{-\frac{1}{2}}\,\chi (x)\nonumber\\
\Psi^\dagger (x) &=& \chi^\dagger (x)\,[1 +
e\not \!\! A (\not \! \partial + m)^{-1} ]^{-\frac{1}{2}} \;.
\label{deft3}
\end{eqnarray}Gerald Dunne (Connecticut), Alex Kovner (Oxford), Bayram Tekin (Oxford)

Concerning the spectrum of the dual theory, we have been able to
relate the induced Chern-Simons coefficient with the statistics of
vortex like excitations, which are expected to be relevant for the
understanding of the physical properties of the dual theory {\cite{kov}}. In
particular, in the case of a dual theory corresponding to spin
one-half massive Dirac fermions, these excitations turn out to obey
Fermi Dirac statistics, provided one fixes the ambiguities of the
induced Chern-Simons coefficient by requiring large gauge invariance
of the fermionic determinant, when the time coordinate is compactified
to a circle $S^1$. We also note that the vortex flux quantization
necessary for the validity of this invariance argument, is strongly
related to the presence of terms in the dual action which depend on
the curvature, as for instance the Maxwell term, and which allow for
finite energy solutions.

Finally, it is interesting to make a parallel with the SU(2) skyrmion
quantization. In this case, Witten \cite{sky} has discussed the
possible spin and statistics of the skyrmions, introducing a
compactified time coordinate on the circle $S^1$. The $SU(2)$ Skyrme
configurations defined on $S^1 \times S^3 $ can be separated into two
distinct homotopy classes, yielding two inequivalent quantizations,
corresponding either to fermions or to bosons. These two possibilities
can be implemented by the introduction of a topological
Wess-Zumino-Witten (WZW) action, which allows to properly distinguish
the two different statistics.  In this sense, the Chern-Simons term of
the dual theory can be interpreted as playing a role analogous to that
of the WZW term in the Skyrme model. Both terms, being of first order
in the time derivative, survive the process where two topological
excitations are adiabatically interchanged.

\section*{Acknowledgements}
The Conselho Nacional de Desenvolvimento Cient\'{\i}fico e
Tecnol\'{o}gico CNPq-Brazil, the Funda{\c {c}}{\~{a}}o de Amparo
{\`{a}} Pesquisa do Estado do Rio de Janeiro (Faperj), and the
SR2-UERJ are acknowledged for the financial support.  C.~D.~F.~would
like to acknowledge the members of the Departament of Theoretical
Physics at the UERJ, for their warm hospitality.  He is partially
supported by ANPCyT (PICT 97/1040) and Fundaci{\'o}n Antorchas grants.


\end{document}